\documentclass[12pt]{revtex4}

\usepackage{graphicx}
\usepackage{dcolumn}
\usepackage{amsmath}

\begin{document}

\title{M$_x$TiSe$_2$ (M = Cr, Mn, Cu) electronic structure study by methods of resonant X-ray photoemission spectroscopy and X-Ray absorption spectroscopy.}

\author{A.S. Shkvarin, Yu.M.~Yarmoshenko}
\affiliation{Institute of Metal Physics Russian Academy of Sciences-Ural Division, 620990 Yekaterinburg, Russia}
\author{M.V.~Yablonskikh}
\affiliation{Sincrotrone Trieste SCpA, Basovizza I-34012, Italy}
\author{N.A.~Skorikov, A.I.~Merentsov, A.N.~Titov}
\affiliation{Institute of Metal Physics Russian Academy of Sciences-Ural Division, 620990 Yekaterinburg, Russia}

\begin{abstract}
Electronic structure and chemical bonding in TiX$_2$ (X=S, Se, Te), TM$_x$TiSe$_2$ (TM=Cr, Mn, Cu) and Cr$_x$Ti$_{1-x}$Se$_2$ were studied by x-ray resonance photoemission and absorption spectroscopy. These methods are detected to be strong sensitive to chemical bonding. Charge transfer from the intercalated atoms to Ti 3d band is detected. Narrow Ti 3d and Cu 3d bands are observed under Fermi level in Cu$_x$TiSe$_2$.
\end{abstract}

\maketitle

\section{Introduction}

\begin {figure}[htb!]
  \centerline{\includegraphics
    [clip=true, width=70 mm]{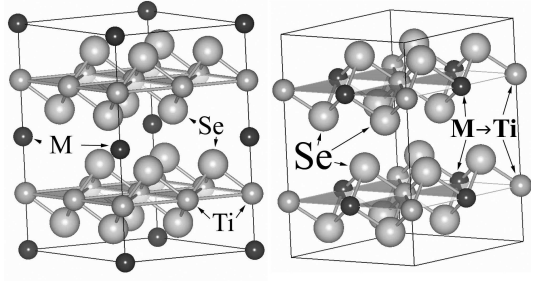}}
\caption {Crystal structures of intercalated TM$_{x}$TiSe$_2$ (left panel) and TM$_{x}$TiSe$_2$ with substitution (right panel)	 are shown.
}
\label {fig1}
\end{figure}

Transition Metals Dichalcogenides (TMDC's) and their intercalates TM$_x$TMDC consist of composite layers of X-Ti-X (X-chalcogen) sticked to each other through weak van der Waals forces interacting along the (0001) plane of crystal. In this paper we discuss the formation of electronic structure of the most interesting representatives of TMDC's: intercalates 1T-TM$_x$TiSe$_2$ (TM = Cr, Cu) and substitution compounds Cr$_{1-x}$Ti$_x$Se$_2$. (Fig. \ref{fig1}).
Single crystals of compounds 1T-Cr$_x$TiSe$_2$, 1T-Cr$_{1-x}$TixSe$_2$ 1T-Mn$_x$TiSe$_2$ and 1T-Cu$_x$TiSe$_2$ with concentration of Cu above of 0.11 mol.\%, were synthesized for the first time. X-ray photoelectron spectra (XPS) of core levels, the resonant valence band spectra (RXPS), and X-ray Ti L$_{2,3}$ absorption spectra (XAS) have been studied. 

In CrSe$_2$ half-metallic density Cr 3d states with large spin polarization \cite{Ref1} was theoretically predicted. Pure CrSe$_2$ cannot be produced in a stable state. This is because the Cr$^{4+}$ ion is unstable. Usually this problem can be solved by introducing additional donor atoms into the lattice \cite{Ref2}. Results of our previous studies had already shown the presence of strong spin-polarization of Cr 3d states for the Cr$_x$TiSe$_2$ \cite{Ref3}. Now we follow another way in our work  -- the substitution of chromium atoms in a regular lattice by the titanium atoms with a higher valence \cite{Ref4}. 

Mn$_x$TiSe$_2$ intercalate for x$<$0.2 show the anomalous properties in comparison with most of TM$_x$TiSe$_2$. This system exhibits under intercalation an increase in the interlayer spacing, growth of conductivity \cite{Ref5}, and anomalously high (by the order of magnitude) increment of the Pauli contribution to the magnetic susceptibility in comparison with the other intercalated metals \cite{Ref5}. 

Cu$_x$TiSe$_2$ exhibit superconducting properties at low concentrations of copper up to 11\% at. \cite{Ref7} and also shows CDW transition. So these materials represent different nature of influence of doping on properties of TiSe$_2$.
 
\section{Results and discussion}

Polycrystalline samples of the 1T-TiS$_2$, TiSe$_2$, TiTe$_2$ and TM$_x$TiSe$_2$ (M = Cr, Cu, Mn) were synthesized from  elements by the ampoule synthesis method. Single crystals were grown by direct evaporation pre-prepared materials in evacuated quartz tubes. The crystals are thin plates about 2$\times$2$\times$0.05 mm$^3$ in size. X-ray photoelectron (XPS), resonant (RXPS) and absorption spectra (XAS) were measured with use of the experimental station MUSTANG in the Russian-German laboratory of BESSY II synchrotron and the standard setup of the BACH and CIPO beamline of synchrotron ELETTRA. All measurements were performed at room temperature. To obtain a fresh surface, the crystals were cleaved in a working chamber at a pressure of 1$\times$10$^{-9}$ mbar.

\begin{table}[!h]
\caption{Binding energy XPS Ti2p$_{3/2}$ levels and photon energy Ti L$_3$ XAS main maximum for TM$_x$TiSe$2$}

\begin{tabular}{ccc}
\label {table1}
 Compound     & Ti2p$_{3/2}$ & Ti L$_3$\\ 
\hline
 TiS$_2$  									& 456.4    & 458\\ 
 TiSe$_2$  									& 455.4    & 457.6\\ 
 TiTe$_2$  									& 454.9    & 456.8\\ 
 Cr$_{0.08}$Ti$_{0.96}$Se$_2$  	& 455.4    & 457.6\\ 
 Cr$_{0.78}$Ti$_{0.36}$Se$_2$  	& 455.6    & 457.6\\ 
 Mn$_{0.1}$TiSe$_2$  					& 455.5    & 456.6\\ 
 Mn$_{0.2}$TiSe$_2$  					& 455.8    & 456.6\\ 
 Cu$_{0.05}$TiSe$_2$  				& 455.4    & 457.6\\ 
 Cu$_{0.6}$TiSe$_2$  					& 455				& 457.6\\ 
\end{tabular}
\end{table}
Atoms of Ti, Cr and Mn are shown to possess the oxidation degree 4+, 3+ and 2+ consequently \cite{Ref3, Ref4, Ref5}. 

The substitution of titanium atoms by chromium in 1T-TiSe$_2$ leads to the formation of structural 1T-CrSe$_2$ fragments whose concentration increases with chromium content \cite{Ref4}. These fragments inherit the crystal structure of the initial phase of 1T-TiSe$_2$ and stabilized by extra-titanium between composite layers. The substitution does not affect the state of the chemical bond of titanium, as suggested by the titanium L$_{2,3}$ absorption spectra with the shape and energy position identical to those of the spectra for initial 1T-TiSe$_2$. (Table \ref{table1}).
The chromium XAS spectra do not change with chromium content. This results from invariability of the local environment of the chromium atom in the structural CrSe$_2$ fragments. 
\begin {figure}[!t]
  \centerline{\includegraphics
    [clip=true, width=60 mm]{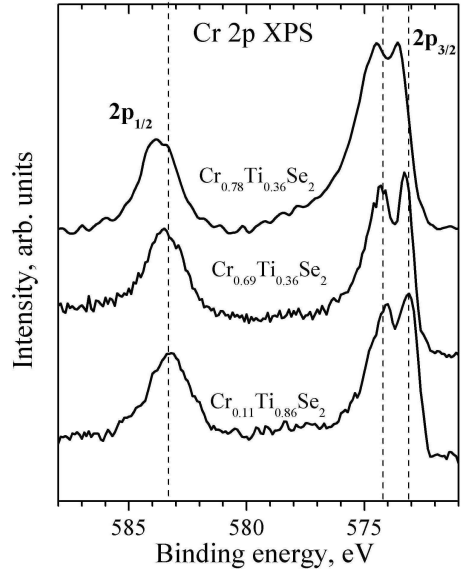}}
\caption {Spectra of the core Cr 2p levels.}
\label {fig2}
\end{figure}
\begin {figure}[!b]
  \centerline{\includegraphics
    [clip=true, width=60 mm]{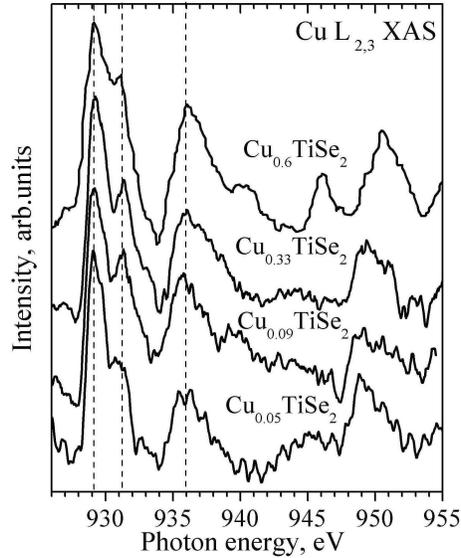}}
\caption {Cu L$_{2,3}$ absorption spectra of Cu$_x$TiSe$_2$ crystals. Spectra with low concentration of Cu are identical to spectra of Cu vapour \cite{Ref12}}
\label {fig3}
\end{figure}An intense narrow Cr 3d band near the Fermi energy was detected during the measuring the RXPS of the chromium valence band \cite{Ref8}. 
\begin {figure*}[!b]
  \centerline{\includegraphics
    [clip=true, width=120 mm]{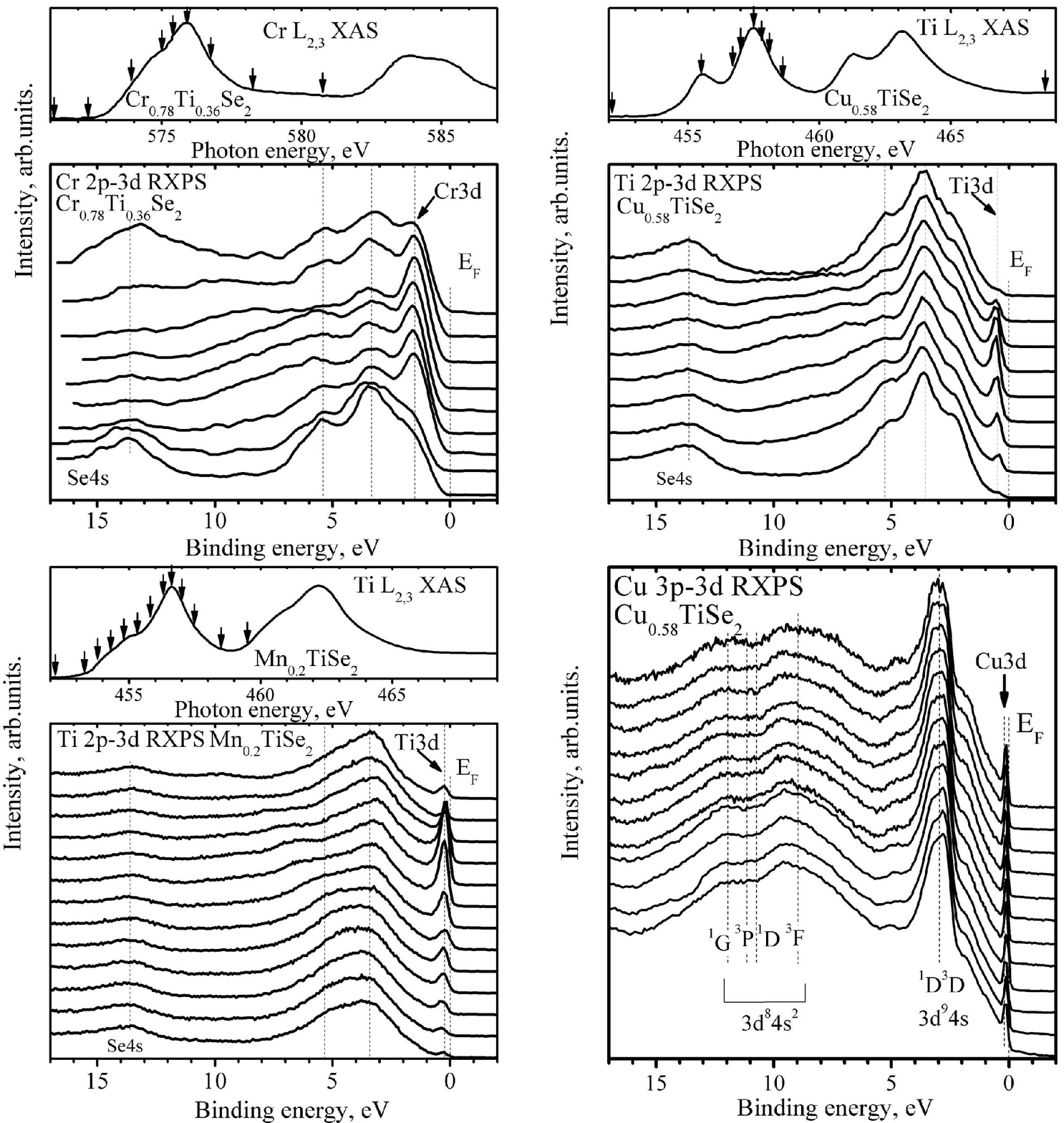}}
\caption {Spectra of the valence bands at resonant excitation. Exitation energy range for Cr 2p-3d resonance is 570-581 eV, for Ti 2p-3d resonance is 452-460 eV, for Cu 3p-3d is 70-81 eV. Arrows at absorption spectra are corresponding to exitation energy. Exitation energy on valence band spectra grows up from bottom to top of figure. In Cu$_{0.58}$TiSe$_2$ valence band spectra exitation energy grows up with step 1 eV.}
\label {fig4}
\end{figure*}
Experimentally detected an energy splitting of the Cr 2p$_{3/2}$ line (Fig. \ref{fig2}), about 1 eV, which is caused by the exchange magnetic interaction between the Cr 3d electrons and the core 2p vacancy \cite{Ref3}. Theoretical calculations predict semimetallic nature Cr3d density of states CrSe$_2$ stabilized by titanium. This result is confirmed by the energy splitting of the core Cr 2p$_{3/2}$ lines.

In the case of Mn the mixed covalent-ionic chemical bonding in the octahedral surrounding identical to that of Ti is observed. These results are confirmed by the atomic multiplet calculations of Ti, Mn XAS and Mn 2p XPS \cite{Ref9}. In the spectrum of Mn 2p XPS \cite{Ref9}, satellites have been registered at a distance of 5 eV from the main lines, which evidences the partly ionic chemical bonding. In XPS of the valence band, on reaching the 2p resonance for Ti, narrow bands localized just below the Fermi level have been detected (Fig. \ref{fig4}). Note that in pure TiSe$_2$ this effect is completely absent. These experimental data indicate the charge transfer from the Mn atoms into the conduction band formed by Ti 3d states. From the theoretical calculations of E(k) it was established that under the Fermi level the Ti 3d states are localized in the vicinity of point $\Gamma$ of the Brillouin zone of the crystal. The Mn 3d states are localized along the direction M-$\Gamma$-K. 

Intercalation of Cu (in contrast to other transition metals) does not change the chemical bonding in the host lattice TiSe$_2$. XAS Ti L$_{2,3}$ having an identical shape and energy position of the pristine 1T-TiSe$_2$ coinfirms this (Table \ref{table1}).
Concentration dependences of unit cell constant and kinetic shows different behavior in Cu content ranges $x$=0-0.5 and 0.5-0.75 \cite{Ref10}. The shape and energy position of the Cu L$_{2,3}$ XAS are not changed in x = 0 - 0.5 range (Fig. \ref{fig3}) with change of $x$.  Analysis of copper XAS shows that the closest analogue in ion classification is Cu$^0$. This conclusion is in agreement with theoretical calculations of the electronic structure. Measurement RXPS Cu 3p shows that the 3d band is almost full. In Cu 3p RXPS detected satellites are corresponding several term in basic configurations Cu d$^8$ \cite{Ref11} (Fig. \ref{fig4}). 
In Cu 3p and Ti 2p RXPS of Cu$_{0.6}$TiSe$_2$ two narrow peaks with E$_B$=0.1 eV and E$_B$=0.3 eV associated with Cu 3d and Ti 3d bands(Fig. \ref{fig4}), respectively, are observed. The appearance of additional bands in the Ti RXPS below the Fermi level means the filling Ti 3d states and indicates the ionic contribution to the chemical bond between the intercalated atoms and TiSe$_2$. In the valence band Cu$_x$TiSe$_2$ below the Fermi level narrow Ti 3d and Cu 3d  bands are separated by an energy gap.
\section{Acknowledgments}
This work was performed within the framework of the "Russian-German Laboratory at BESSY" bilateral Program, project N 81041. Work supported by Russian foundation for basic research, grant N 09-03-00053, CRDF RUX0-000005-EK-06/BP4M05 and Program of Interdisciplinary research of UrD RAS, project N 36. Work on synchrotron ELETTRA supported by international centre fur theoretical physics (ITCP), projects N 20100397 and N 20105305.


\begin{thebibliography}{99}
\bibitem{Ref1} C M Fang et al. J. Phys.: Condens. Matter 9 (1997) 10173
\bibitem{Ref2}  J. Dijkstra et al. Phys. Rev. B 40 (1989), 7973 
\bibitem{Ref3} A.N.~Titov et al. Phys. Rev. B 63 (2001) 035106
\bibitem{Ref4}A.I.Merentsov et al. J. Electron Spectrosc. Relat. Phenom. 182 (2010) 70
\bibitem{Ref5} V.I.~Maksimov et al.  Journal of Alloys and Compounds 384 (2004) 33-38
\bibitem{Ref6} Y.~Tazuke et al. J. Phys. Soc. Japan 66 (1977) 827
\bibitem{Ref7} E.~Morosan et al. Nature Physics 2 (2006) 544 - 550
\bibitem{Ref8} A.S.Shkvarin et al.  JETP 112 (2011) 87
\bibitem{Ref12} A.~Verweyen et al. J. Phys. B: At. Mol. Opt. Phys. 33 (2000) 1563–1573
\bibitem{Ref9} M.V.Yablonskikh et al.  arXiv (2011) vol. cond-mat.mtrl-sci, http://arxiv.org/abs/1101.2160v1
\bibitem{Ref10} A.A.Titov et al. Phys. Solid State 51, (2009) 217
\bibitem{Ref11} M.~Iwan et al. Phys. Rev. lett. 43 (1979) 1829–1832
\end{thebibliography}
\end{document}